\documentclass[10pt,showpacs,longbibliography,amsmath,amssymb,osajnl,floatfix,superscriptaddress,twocolumn]{revtex4-2}
\usepackage{graphicx}
\usepackage{dcolumn}
\usepackage{bm}
\usepackage{color}
\usepackage{txfonts}
\usepackage{microtype}
\usepackage{amsmath}
\usepackage{slashed}
\usepackage{comment}
\usepackage{ulem}

\begin{document}

\title{Brilliant circularly polarized $\gamma$-ray sources via single-shot laser plasma interaction}

\author{Yu Wang}  \thanks{These authors have contributed equally to this work.}
\affiliation{Ministry of Education Key Laboratory for Nonequilibrium Synthesis and Modulation of Condensed Matter, Shaanxi Province Key Laboratory of Quantum Information and Quantum Optoelectronic Devices, School of Physics, Xi'an Jiaotong University, Xi'an 710049, China}
\author{Mamutjan Ababekri}  \thanks{These authors have contributed equally to this work.}
\affiliation{Ministry of Education Key Laboratory for Nonequilibrium Synthesis and Modulation of Condensed Matter, Shaanxi Province Key Laboratory of Quantum Information and Quantum Optoelectronic Devices, School of Physics, Xi'an Jiaotong University, Xi'an 710049, China}
\author{Feng Wan} \email{wanfeng@xjtu.edu.cn}
\affiliation{Ministry of Education Key Laboratory for Nonequilibrium Synthesis and Modulation of Condensed Matter, Shaanxi Province Key Laboratory of Quantum Information and Quantum Optoelectronic Devices, School of Physics, Xi'an Jiaotong University, Xi'an 710049, China}
\author{Qian Zhao}
\affiliation{Ministry of Education Key Laboratory for Nonequilibrium Synthesis and Modulation of Condensed Matter, Shaanxi Province Key Laboratory of Quantum Information and Quantum Optoelectronic Devices, School of Physics, Xi'an Jiaotong University, Xi'an 710049, China}
\author{Chong Lv}
\affiliation{Department of Nuclear Physics, China Institute of Atomic Energy, P. O. Box 275(7), Beijing 102413, China}
\author{Xue-Guang Ren}
\affiliation{Ministry of Education Key Laboratory for Nonequilibrium Synthesis and Modulation of Condensed Matter, Shaanxi Province Key Laboratory of Quantum Information and Quantum Optoelectronic Devices, School of Physics, Xi'an Jiaotong University, Xi'an 710049, China}
\author{Zhong-Feng Xu}
\affiliation{Ministry of Education Key Laboratory for Nonequilibrium Synthesis and Modulation of Condensed Matter, Shaanxi Province Key Laboratory of Quantum Information and Quantum Optoelectronic Devices, School of Physics, Xi'an Jiaotong University, Xi'an 710049, China}
\author{Yong-Tao Zhao}
\affiliation{Ministry of Education Key Laboratory for Nonequilibrium Synthesis and Modulation of Condensed Matter, Shaanxi Province Key Laboratory of Quantum Information and Quantum Optoelectronic Devices, School of Physics, Xi'an Jiaotong University, Xi'an 710049, China}
\author{Jian-Xing Li}
\affiliation{Ministry of Education Key Laboratory for Nonequilibrium Synthesis and Modulation of Condensed Matter, Shaanxi Province Key Laboratory of Quantum Information and Quantum Optoelectronic Devices, School of Physics, Xi'an Jiaotong University, Xi'an 710049, China}


\begin{abstract}
Circularly polarized (CP) $\gamma$-ray sources are versatile for broad applications in nuclear physics, high-energy physics and astrophysics. The laser-plasma based particle accelerators provide accessibility for much higher flux $\gamma$-ray sources than conventional approaches, in which, however, the circular polarization properties of emitted $\gamma$-photons are used to be neglected. In this letter, we show that brilliant CP $\gamma$-ray beams can be generated via the combination of laser plasma wakefield acceleration and plasma mirror techniques. 
In weakly nonlinear Compton scattering scheme with moderate laser intensities, the helicity of the driving laser can be transferred to the emitted $\gamma$-photons, and their average polarization degree  can reach about $\sim 37\%$ ($21\%$) with a peak brilliance of $\gtrsim 10^{21}~$photons/(s $\cdot$ mm$^2 \cdot$ mrad$^2 \cdot$ 0.1\% BW) around 1~MeV (100~MeV). Moreover, our proposed method is easily feasible and robust with respect to the laser and plasma parameters. 
\end{abstract}
\date{\today}
\maketitle

The polarization property of the $\gamma$-rays is of great significance to reveal the emission mechanisms in the pulsar \cite{Dean2008,Buhler2013}, magetars and other galactic objects \cite{Orlando2013}. As an essential tool, $\gamma$-ray beams can be used in researches of nuclear physics \cite{Weller2003,Weller2009}, high-energy physics \cite{Schoeffel2020}, industrial applications of medical imaging \cite{Assmann2020} and object tomography, etc. In particular, circularly polarized (CP) ones have broad applications in, for instances, the generation of longitudinally-polarized positron beams \cite{Omori2006, Moortgat2008}, polarization-dependent photo-fission of nucleus in the giant dipole resonance \cite{Speth1981} and photo-production of mesons \cite{Li2020}. Conventionally, highly polarized photons can be generated via free electron laser (FEL), Thomson and Compton scattering, bremsstrahlung and synchrotron radiation \cite{Schaerf2005}. However, FEL can only deliver X-ray photons of tens of keV, and the peak brilliance and flux of other high-energy $\gamma$-ray sources  are normally limited by the scattering probabilities and flux of the driving electron beam. For instance, the peak flux of state-of-the-art High Intensity Gamma-ray Source (HI$\gamma$S) \cite{Weller2009} is about $10^{10}$~photons$\cdot\mathrm{s}^{-1}$ around 10~MeV with beam diameter of  $D\simeq12$~mm. 
Recently, rapid developments of ultra-short ultra-intense laser technologies \cite{Danson2019,Yoon2021} have promoted  high-brilliance laser-plasma based particle \cite{Esarey2009,Macchi2013} and radiation sources \cite{Corde2013,Sarri2014,Yan2017}. When multi-PW  lasers interacting with plasmas, MeV-GeV $\gamma$-photons can be produced with ultrahigh brilliance of $10^{25}$-$10^{27}~\mathrm{photons/(s \cdot mm^2 \cdot mrad^2 \cdot 0.1\% BW)}$ \cite{Ridgers2012,Zhu2020,Xue2020}. However, these produced $\gamma$-photons are either unpolarized or only linearly polarized. And all-optical CP $\gamma$-ray beams can be generated via  multi-PW laser pulses colliding with high-energy longitudinally spin-polarized electron beams in strongly nonlinear Compton scattering \cite{Li2020,King2020}, which faces potential difficulties of spatial and temporal synchronization. Thus, efficient generation of brilliant CP $\gamma$-ray beam is still a great challenge.

\begin{figure}[!h]
	\begin{center}
	\includegraphics[width=\linewidth]{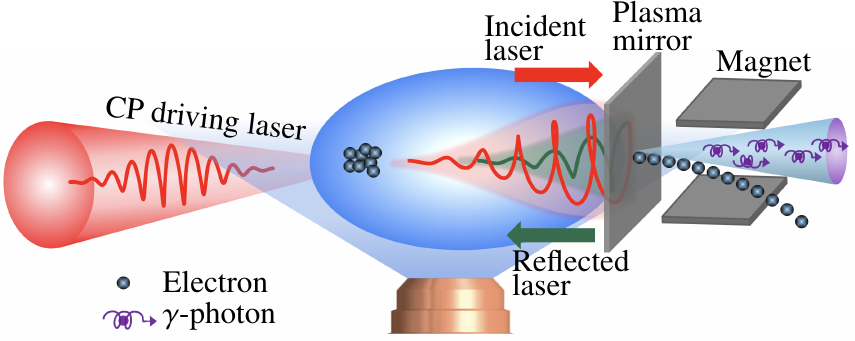}
		\caption{Interaction scenario for  all-optical generation of brilliant CP $\gamma$-ray beam via single-shot laser plasma interaction.}\label{fig1}
	\end{center}
\end{figure}

\begin{figure}[t] 
	\begin{center}
		\includegraphics[width=\linewidth]{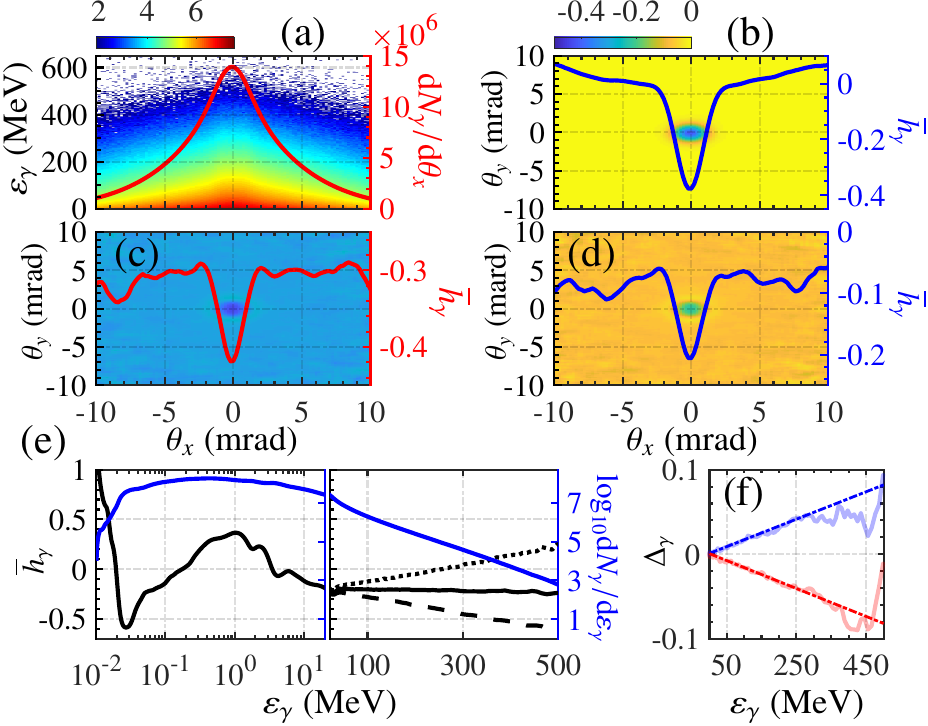}
		\caption{
			(a) Number density of emitted photons d$^2N_\gamma/$d$\theta_x$d$\varepsilon_{\gamma}$ (mrad$^{-1}$~MeV$^{-1}$) vs deflection angle $\theta_x=p_{\gamma,x}/p_{\gamma,z}$ and $\gamma$-photon energy $\varepsilon_\gamma$. (b)-(d) Angle-resolved average helicity $\overline{h}_\gamma$ of emitted photons with unpolarized electrons (electron helicity $h_e = 0$), and, polarized electrons with $h_e =-1$ and $+1$, respectively. In (c) and (d), only photons with $\varepsilon_{\gamma} \geq 100~$MeV are taken into account. Solid lines indicate the lineout along the deflection angle $\theta_y = p_{\gamma,y}/p_{\gamma,z} = 0$ with $\Delta\theta_y = 0.12~$mrad. (e) Blue lines indicate the emission spectra d$N_\gamma/$d$\varepsilon_\gamma~$(MeV$^{-1}$) vs $\varepsilon_\gamma$, and black solid, dashed and dotted lines  indicate $\overline{h}_\gamma$ vs $\varepsilon_{\gamma}$ for the cases of $h_e = 0$, $-1$ and $+1$, respectively. (f) Relative deviation of energy spectra $\Delta_\gamma = 2 \frac{\mathcal{N}_\mathrm{pol}-\mathcal{N}_\mathrm{unpol}}{\overline{\mathcal{N}_\mathrm{pol}+\mathcal{N}_\mathrm{unpol}}}$ with $\mathcal{N} \equiv~$d$N_\gamma/$d$\varepsilon_\gamma$, where blue and red lines indicate $h_e = +1$ and $-1$, respectively; solid lines are from the simulation results, and dash-dotted lines are fitting curves.}\label{fig2}
	\end{center}
\end{figure}

\begin{figure}[t]
	\begin{center}
		\includegraphics[width=\linewidth]{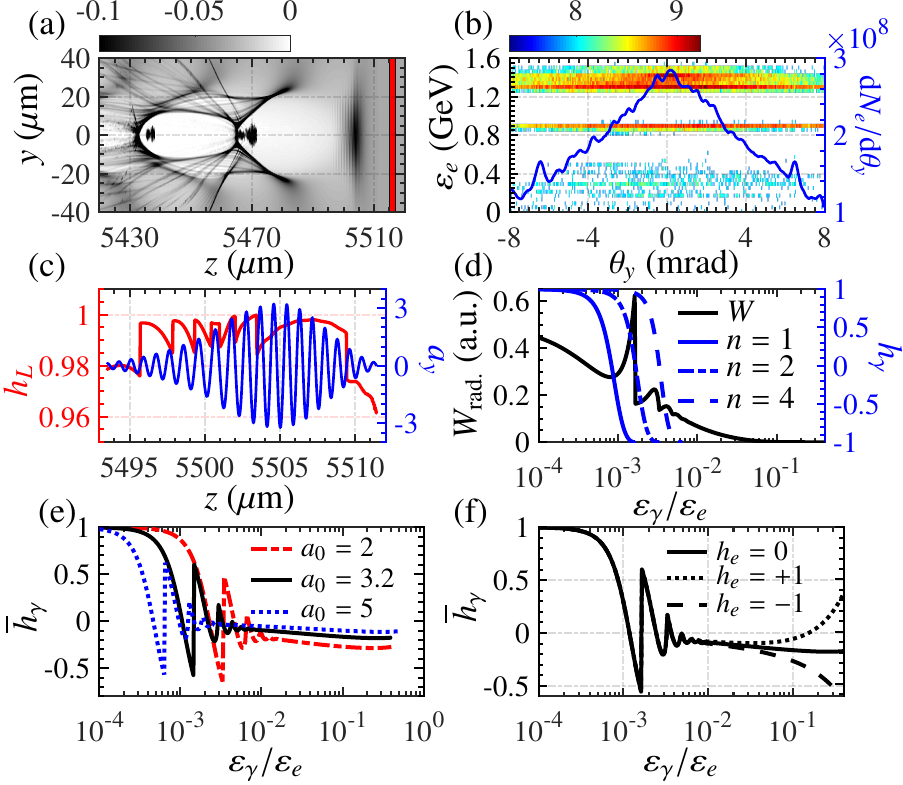}
		\caption{
			(a) Number density of accelerated electrons $n_e/n_c$ in $z$-$y$ plane with $x=0$. The red band denotes the aluminum plasma target. (b) Number distribution of electrons $\log_{10}[$d$^2N_e/$d$\theta_x$d$\varepsilon_e$]~(mrad$^{-1}$~MeV$^{-1}$) vs $\theta_x$ and  $\varepsilon_{e}$.  (c) Blue and red lines indicate the  envelop and  helicity of the reflected laser pulse, respectively. (d) Black solid line indicates the emission rate [normalized by $W_0$ in Eq.~(\ref{eqn1})] with respect to  $\varepsilon_{\gamma}/\varepsilon_{e}$. Blue solid, dash-dotted and dashed lines indicate $h_\gamma$ as absorbing 1, 2 and 4 photons, respectively. (e) and (f) $\overline{h}_\gamma$ vs  $\varepsilon_\gamma/\varepsilon_e$ for different laser intensities and electron helicities, respectively.}\label{fig3}
	\end{center}
\end{figure}

\begin{figure}[t]
	\begin{center}
	\includegraphics[width=\linewidth]{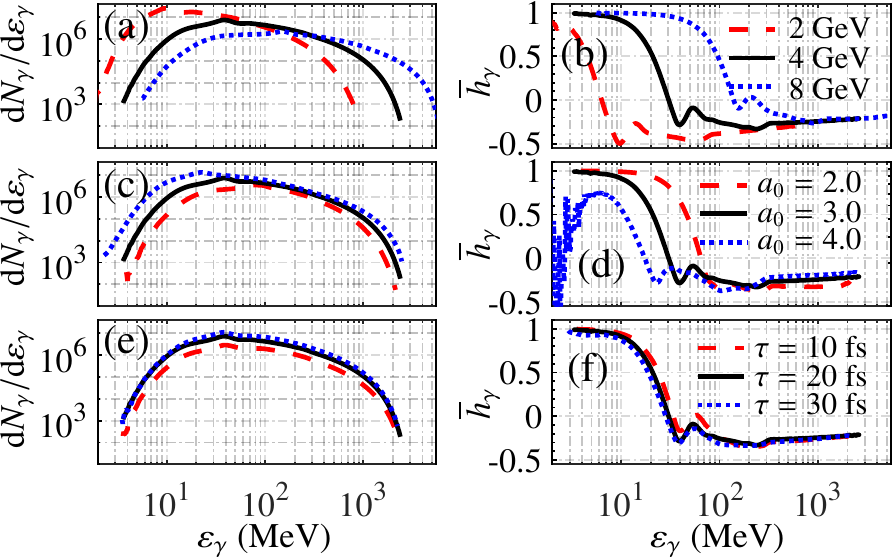}
		\caption{Impact of  [(a) and (b)] $\varepsilon_{e,\mathrm{peak}}$, [(c) and (d)] $a_0$, [(e) and (f)] laser pulse duration $\tau$ on $\overline{h}_\gamma$. Other parameters are the follows: for the electrons $\varepsilon_{e,\mathrm{peak}} = 4$~GeV, energy spread $3\%$, $\Delta\theta_e = 0.2$~mrad, transverse beam size $\sigma_t = 2.1~\mu\mathrm{m}$, beam length $l_e = 5~\mu\mathrm{m}$ and total charge 300~pC; for the driving laser,  the spatial profile is Gaussian with $a_0 = 3.0$, $\tau = 20~\mathrm{fs}$ and transverse focal radius $w_0 = 11~\mu \mathrm{m}$.}\label{fig4}
\end{center}
\end{figure}

In this Letter, we put forward an efficient all-optical method for generating brilliant CP $\gamma$-ray beams via single-shot laser plasma interaction. See the interaction scenario in Fig.~\ref{fig1}, which is similar to the setup in Ref.~\cite{Phuoc2012}, however, here we take into account the helicity transfer  in the laser plasma interaction. A moderately intense CP laser incidents into a gas plasma and drives the laser plasma wakefield acceleration (LWFA) to generate high-energy electrons. The CP driving laser pulse is then reflected by the plasma mirror and further collides with accelerated electrons to emit CP $\gamma$-photons via weakly nonlinear Compton scattering ($e+n\omega_L \rightarrow e'+\omega_\gamma$). 
Here we find that the helicities of emitted $\gamma$-photons in the low-energy regime is mainly associated with those of the driving laser photons. While, for high-energy $\gamma$-photons, due to  multi-photon absorption, the average helicity will saturate to a constant for unpolarized eletrons or linearly depend on the energies of emitted photons for polarized electrons; see more details in Fig.~\ref{fig3}. (By contrast, in strongly nonlinear Compton scattering the circular polarization of emitted $\gamma$-photon is determined by the electron helicity \cite{Li2020}.) 
With a moderately intense CP laser, CP $\gamma$-photon beams with brilliance of $\gtrsim 10^{21}~\mathrm{photons/(s \cdot mm^2 \cdot mrad^2 \cdot 0.1\% BW)}$ can be generated with polarization degree of $\simeq 37\%$; see more details in Fig.~\ref{fig2}.
Moreover, the proposed method is robust with respect to the laser and plasma parameters; see more details in Fig.~\ref{fig4}.

In our simulations, we use three-dimensional (3D) particle-in-cell (PIC) code EPOCH \cite{Arber2015} to simulate the LWFA process, and the Monte Carlo code CAIN \cite{CAIN242,Li2020} to simulate the weakly nonlinear Compton scattering process. 
As an example, we employ a right-hand CP laser pulse (helicity $h_L = -1$) propagating along $z$ direction with wavelength $\lambda_0 = 0.8~\mu\mathrm{m}$, normalized intensity $a_0 \equiv e E_\mathrm{rms} /m_ec\omega_L = 2.8$ [corresponding peak intensity $I_0 \approx 2.74\times10^{18} a_0^2 \left(1~\mu\mathrm{m}/\lambda_0\right)^2~\mathrm{W/cm^2} \approx 3.46 \times 10^{19}~\mathrm{W/cm^2}$], and transverse Gaussian profile with focal radius $w_0 = 11.5~\mu\mathrm{m}$, where $e$ and $m_e$ are the charge and mass of the electron, respectively, $E_\mathrm{rms} \equiv \langle (E^2)^{1/2} \rangle$, $E$ and $\omega_L$ the root mean square (rms) electric field, electric field and frequency of the laser field, respectively, and $c$ the light speed in vacuum.
The temporal profile is composed by a flat-top part of $\tau_\mathrm{flat}=6T_0$ and rising (falling) part of $\tau_\mathrm{rising~(falling)} = 12T_0$ with Gaussian-like 5th order symmetric polynomial $10t'^3 - 15t'^4 + 6t'^5$ (here, $t'\equiv t/T_0$ and $T_0$ is the laser period) \cite{Lu2007}. 
The number density of the gas plasma is linearly rising from $n_e = 0$ at $z=0$ to $n_e = 1.5 \times 10^{18}~\mathrm{cm^{-3}}$ at $z=100~\mu\mathrm{m}$ and then distributed uniformly to $z=5.515~$mm. The aluminum plasma mirror is placed at $z=5.515~$mm with a thickness of $l=2~\mu\mathrm{m}$, number density $n_e = 451n_c$ and scale length $L = \lambda_0$ for the preplasma \cite{Esirkepov2014}, where the critical plasma density is $n_c = m_e\omega_L^2 / 4\pi e^2$. 
The spatial sizes of the simulation box are $z \times x \times y = 127\lambda_0 \times 80\lambda_0 \times  80\lambda_0$ with cell sizes of $4000 \times 256 \times 256$. 
The numbers of macro-particle per cell are assigned as 2 and 1 for electrons and Helium ions, respectively.
Note that the above mentioned spatial sizes are not fine enough to resolve the plasma frequency of the aluminum target, thus the reflection of the laser pulse is recalculated with finer grid sizes of $\Delta z \times \Delta x = \frac{\lambda_0}{200} \times \frac{\lambda_0}{50}$ via the two dimensional PIC to efficiently reduce the massive 3D computation time, and, the  numbers of macro-particle per cell for electrons and ions are set to 100 and 30, respectively.

Simulation results of emitted CP $\gamma$-ray beam  are shown in Fig.~\ref{fig2}. 
The peak intensity of emitted $\gamma$-rays is in the order of $10^7~\mathrm{mrad^{-1}}$ for photon energies $\varepsilon_\gamma \geq 5~\mathrm{keV}$. 
The corresponding brilliances (average helicities $\overline{h}_\gamma$) are $4.35 \times 10^{21}~(0.37)$, $1.37 \times 10^{22}~(-0.11)$, $5.89 \times 10^{21}~(-0.20)$ and  $1.46 \times 10^{21}~(-0.21)~\mathrm{photons/(s \cdot mm^2 \cdot mrad^2 \cdot 0.1\% BW)}$ for $\varepsilon_{\gamma} =~$1~MeV, 10~MeV, 100~MeV and 200~MeV, respectively.  
Here, the angular spread of the $\gamma$-ray beam $\Delta\theta_\gamma$ originates from the angular spread of the electron beam $\Delta\theta_e$ and the laser induced transverse momentum $p_{e,\perp}=\sqrt{p_{e,x}^2+p_{e,y}^2} \sim a_0$, i.e., $\Delta\theta_{\gamma} \simeq \Delta\theta_{e} + 2p_{e,\perp}/p_{e,z} \simeq \Delta \theta_{e} + 2 a_0/\overline{\gamma}_e$, where $\overline{\gamma}_e$ is the average Lorentz factor of electrons. Note that here for ultra-relativistic electrons $\gamma_e\gg 1$ the $\gamma$-photons are assumed to be emitted along the electron momentum (due to the emission solid angle $\sim 1/\gamma_e$).  
However, since high-energy electrons (with electron energies $\varepsilon_{e} \approx 900~$MeV and 1400~MeV) are concentrated in a narrow angle of less than 1~mrad,  $\Delta\theta_\gamma \approx 2 a_0 / p_{e,z} \approx 3$-4~mrad [see  Figs.~\ref{fig2}(a) and \ref{fig3}(b)].
The average helicity $\overline{h}_\gamma$ of $\gamma$-photons within a narrow cone of  $\Delta\theta\lesssim 1~$mrad can reach $-0.38$, but for other photons of $\Delta\theta \gtrsim 3~$mrad, $\overline{h}_\gamma$ is in the range of ($-0.01$, $0.01$) [see Fig.~\ref{fig2}(b)].
The energy-resolved $\overline{h}_\gamma$ presents quite different feature, and can reach nearly 1.0 at low energies of $\sim$10~keV.
$\overline{h}_\gamma$ quickly drops to $\sim -0.59$ in the vicinity of 20~keV and then rises to $\sim 0.37$ around 1~MeV [see Fig.~\ref{fig2}(e)].
In the high-energy regime $\varepsilon_\gamma \gtrsim 20$~MeV, $\overline{h}_\gamma$ saturates at $\simeq -0.2$.
When employing polarized electron bunches, the radiation spectra and angular distribution are almost identical to the unpolarized case with an relative deviation of 0.9\%, 1.7\% and 3.3\% around $\varepsilon_\gamma \approx 50, 100$ and 200 MeV, respectively [see Fig.~\ref{fig2}(f)].
Angle-resolved $\overline{h}_\gamma$ of polarized cases also show similar patterns to that in  Fig.~\ref{fig2}(b) and consequently are excluded.
For $\varepsilon_\gamma \gtrsim 100$~MeV,  angle-resolved $\overline{h}_\gamma$ is in the order of $-0.31~(-0.08)$ for $h_e = -1~(+1)$ [see Figs.~\ref{fig3}(c) and (d)].
Especially, in a narrow cone of $\Delta\theta \leq 1~$mrad, $\overline{h}_\gamma \simeq -0.42~(-0.21)$ for $h_e=-1(+1)$ which means that large numbers of $\gamma$-photons with $\varepsilon_{\gamma} \simeq (20$-$40)~$keV are generated with small angular spread [see Figs.~\ref{fig2}(c)-(e)].  
Besides, for $\gamma$-photons with energies of $\varepsilon_\gamma \lesssim 20$~MeV, $\overline{h}_\gamma$ is identical to the unpolarized case.
However, in the high-energy part of $\varepsilon_\gamma \gtrsim 20$~MeV, $\overline{h}_\gamma$ is linearly rising (falling) as $\varepsilon_\gamma$ in the case of $h_e = +1~(-1$) and reaches $\simeq 0.25~(-0.61)$ at the energy cutoff of $\varepsilon_\gamma \simeq 500$~MeV [see Fig.~\ref{fig2}(e)].
Such brilliant CP $\gamma$-ray beam can be used for polarized lepton creation with $\varepsilon_\gamma \gtrsim 1.022~\mathrm{MeV}~(2m_ec^2)$ \cite{Moortgat2008} and photo-nuclear physics with $\varepsilon_\gamma \simeq 10$-$100$~MeV~\cite{Speth1981}.  

The physical reasons of generating the CP $\gamma$-ray beam are analyzed in Fig.~\ref{fig3}. 
In LWFA process, two plasma bubbles are excited and trap electrons [see Fig.~\ref{fig3}(a)] to create two isolated quasi-monoenergtic electron bunches with peak energy $\varepsilon_{e,\mathrm{peak}}\approx$ 900~MeV and 1.4~GeV, respectively [see Fig.~\ref{fig3}(b)] (similar to those in \cite{Lu2007}). 
The total numbers of accelerated electrons are $4.6\times10^{8}$ (74~pC) with energy spread of $5\%$ at $\varepsilon_{e,\mathrm{peak}}\approx$ 900~MeV and $1.8\times 10^9$ (288~pC) with energy spread of $10\%$ at $\varepsilon_{e,\mathrm{peak}}\approx$ 1.4~GeV, respectively. 
Angular spread of all electrons is $\Delta \theta_e \approx 12~\mathrm{mard}$, but for high-energy electrons near $\varepsilon_{e,\mathrm{peak}} \simeq 900~$MeV (1.4~GeV), $\Delta \theta_e \simeq 1~$mrad  [see Fig.~\ref{fig3}(b)]. 
Due to the inhomogeneity of the electron density, the driving laser is chirped when propagating in the front of the wakefield [see the $y$-component of the reflected laser in Fig.~\ref{fig3}(c)]. 
After reflection, due to the frequency chirping, the helicity of the driving laser is flipped and changes from negative $h_L=-1$ (right-hand rotation) to positive $\sim 0.98$ (left-hand rotation) (the laser helicity is calculated via $h_L = 2 \mathrm{Im}(E_x^*E_y)/|E_x^*E_y|$ \cite{Wiley1991, CAIN242}, where $E_{x,y}$ are the complex amplitudes of the electric field in $x$ and $y$ directions, respectively, and $E_x^*$ is the complex conjugate of $E_x$). 
Note that the deviation due to the frequency chirping is evaluated via semi-classical calculations~\cite{Wistisen2019} and the average relative errors  in energy spectra and helicity are both only about $1.3\%$. 

When electrons scatter with the reflected CP laser, they may absorb single or multiple low-energy laser photons and then emit a high-energy $\gamma$-photon via nonlinear Compton scattering.  In the weakly nonlinear regime ($a_0\gtrsim 1$), the polarization-dependent cross section is given by~\cite{Tsai1993,CAIN242}
\begin{equation}
	W_{if} = W_0\sum_{n=1}^{\infty} \int_{0}^{\delta_n} d\delta[F_{1n}+h_L h_e F_{2n}+h_\gamma (h_L F_{3n}+h_e F_{4n})], \label{eqn1}
\end{equation}
with the photon helicity 
\begin{equation}
	h_\gamma = \frac{h_LF_{3n}+h_eF_{4n}}{F_{1n}+h_Lh_eF_{2n}}, \label{eqn2}
\end{equation}
where $W_0$ = $\frac{\alpha m_e^2 a_0^2}{4\varepsilon_\mathrm{eff}}$, $\alpha=1/137$ is the fine structure constant, $\varepsilon_\mathrm{eff} = \varepsilon_e + a_0^2{\varepsilon}_L/\Lambda $ the effective energy of initial electron in the laser field, $\delta = (k_{\gamma} \cdot k_L)/(k_L \cdot p)$, $\delta_n = n\Lambda/(1+a_0^2+n\Lambda)$ the cutoff energy fraction of emitted photon absorbing $n$ laser photons \cite{Tsai1993}, $\Lambda = 2(k_L \cdot p)/m_e^2$ the laser energy parameter, $p$, $k_L$ and $k_\gamma$ the four-momenta of the initial electron, laser photon and emitted photon, respectively, $\varepsilon_L$ the energy of the laser photon. $F_{kn}$ ($k = 1,2,3,4$) in Eq.~(\ref{eqn1}) are given in detail in Refs.~\cite{Tsai1993, Ivanov2004}. In strongly nonlinear Compton scattering with $a_0 \gg 1$, the photon polarization $h_\gamma$ is mainly determined by the electron helicity $h_e$ \cite{Baier1998,Li2020}, however, in the weakly nonlinear regime, $h_\gamma$ not only depends on the electron helicity $h_e$, but also on the scattering laser helicity  $h_L$; see Eq.~(\ref{eqn2}). 

For unpolarized electrons ($h_e = 0$), the average helicity of  emitted $\gamma$-photons via the $n$-photon absorption channel is given by $\overline{h}_\gamma = \frac{\sum_n F_{3n}}{\sum_n F_{1n}}h_L$. 
As $\varepsilon_\gamma/\varepsilon_e \sim \delta \ll \delta_1 \approx 0.0015$, the emitted photons are mainly contributed by the one-photon absorption channel ($n=1$) and $\overline{h}_\gamma \approx 1$, i.e., the helicities of emitted $\gamma$-photons are solely determined by the laser helicity [see Fig.~\ref{fig3}(e)]. 
For $10^{-3} \lesssim \delta \lesssim 7\times10^{-3}$, corresponding to multi-photon absorption with $1 \leq n \leq 5$, $\overline{h}_\gamma$ is rapidly oscillating due to the competition among different multi-photon absorption channels with significant gaps of ($\delta_n-\delta_{n-1}$) for small $n$ [see $h_\gamma$ for different channels in Fig.~\ref{fig3}(d) and average $\overline{h}_\gamma$ in Fig.~\ref{fig3}(e)].
For $10^{-2} \lesssim \delta \lesssim 0.4$, $\overline{h}_\gamma$ saturates to $\sim -0.15$ [see Fig.~\ref{fig3}(e)].  
Above theoretical analysis further confirms our simulation results in
 Fig.~\ref{fig2}. For instance, in the low-energy part ($\varepsilon_{\gamma} \lesssim 10~$keV) with $\delta \simeq 1\times 10^{-5} \ll \delta_1$, one obtains $\overline{h}_\gamma \approx 1$, while in the high-energy part ($\varepsilon_{\gamma} \gtrsim 20~$MeV) with $\delta \gtrsim 0.014$-$0.022 \gtrsim \delta_5$,  $\overline{h}_\gamma$ saturates to $\sim -0.21$ [see Fig.~\ref{fig2}(e)]. Moreover, since the first Compton edge (i.e., the cutoff energy of one-photon absorption channel) occurs at $\varepsilon_{\gamma} \simeq 1.3$-$2$~MeV, $\overline{h}_\gamma$ peaks around $\varepsilon_{\gamma}\approx 1$~MeV, and, due to the mixture of $\overline{h}_\gamma$ derived from $\varepsilon_{e}\approx 900$~MeV and 1.4~GeV, the valley zone of $\overline{h}_\gamma$ near $10^{-4} \lesssim \delta \lesssim \delta_1$ (10~keV $\lesssim \varepsilon_{\gamma} \lesssim$ 1~MeV) is broadened and the fast oscillation near $10^{-3} \lesssim \delta \lesssim 10^{-2}$ (1.8~MeV $\lesssim \varepsilon_{\gamma} \lesssim$ 15~MeV) is smoothed [see the comparison between Fig.~\ref{fig2}(e) and Fig.~\ref{fig3}(e)].   
Note that in the high-energy part ($\varepsilon_\gamma \gtrsim 20~$MeV) in Fig.~\ref{fig3}(e), $\overline{h}_\gamma$ saturates to $\sim -0.21$ which lies between the analytical saturation values of $-0.18$ and $-0.28$ for $a_0 = 3.2$ and 2.0 [see Fig.~\ref{fig2}(e) and Fig.~\ref{fig3}(e)], and the slight deviation is derived from  the finite-pulse effect in our numerical simulation (by comparison, we employ a monochromatic plane wave in the analytical estimation).

Furthermore, when electrons are polarized~\cite{Wen2019,Nie2021}, the impact of the electron helicity $h_e$ on the emitted photon helicity $\overline{h}_\gamma$ is negligible in the low-energy  regime of $\varepsilon_\gamma/\varepsilon_e \lesssim 10^{-2}$ [see Fig.~\ref{fig2}(e) and Fig.~\ref{fig3}(f)].  
However, as $\varepsilon_\gamma / \varepsilon_e \gtrsim 10^{-2}$, $\overline{h}_\gamma$ will linearly increase (decrease) for $h_e=+1~(-1)$ [see Fig.~\ref{fig3}(f)]. 
This analytical tendency completely coincides with our numerical results in  Fig.~\ref{fig2}(e).

For the experimental feasibility, the impact of the laser and plasma parameters on $\overline{h}_\gamma$ is analyzed in Fig.~\ref{fig4}. 
In the LWFA process, the peak energy of accelerated electrons $\varepsilon_{e, \mathrm{peak}}$ scales with the laser power $P$ and plasma density $n_e$ via $\varepsilon_e \propto P^{1/3}n_e^{-2/3}\lambda_0^{-4/3}$ \cite{Lu2007}.
Thus, in Figs.~\ref{fig4}(a) and (b), the impact of $n_e$ is simulated via $\varepsilon_e \propto n_e^{-2/3}$ with $\varepsilon_{e, \mathrm{peak}} = 2,4$, and 8~GeV corresponding to $n_e = 8.8\times10^{17}$, $3.1\times10^{17}$ and $1.1\times10^{17}~\mathrm{cm^{-3}}$, respectively. As the emission rate $W_0 \propto 1/\varepsilon_{e, \mathrm{eff.}} \propto 1/\varepsilon_e$ [see Eq.~(\ref{eqn1})], lower  (higher) energies of incident electrons will induce higher (lower) yields [see Fig.~\ref{fig4}(a)]. Since the $n$-photon absorption cutoff $\varepsilon_{\gamma,\mathrm{cutoff}} = \delta_n \varepsilon_{e}$ will shift due to the variation of electron energy (but the shift in $\delta_n$ is quite small), the energy spectra in Fig.~\ref{fig4}(a) and $\overline{h}_\gamma$ in Fig.~\ref{fig4}(b) will shift to left (right) as increasing (decreasing) the electron energy. But, the final saturation $\overline{h}_\gamma \simeq -0.2$ remains unchanged [see Fig.~\ref{fig4}(b)]. 
The laser intensity $a_0$ will affect both the emission probability and the nonlinearity of scattering process. As $W_0 \propto a_0^2$, the radiation intensity will increase (decrease) for smaller (larger) $a_0$ [see  Fig.~\ref{fig4}(c)]. For larger $a_0$,  $\delta_n$ will be smaller, thus, $\overline{h}_\gamma$ will shift to left, and vice versa [see Fig.~\ref{fig4}(d)]. 
In addition, increasing (decreasing) the laser pulse $\tau$ will yield more (fewer) photons per electron [see Fig.~\ref{fig4}(e)]. However, due to short pulse effect (average $a_0$ will be lower for shorter pulse), $|\overline{h}_\gamma|$ is slight higher for $\tau=10$~fs than $\tau=20$~fs and 30~fs [see Fig.~\ref{fig4}(f)].

In conclusion, we put forward an efficient brilliant CP $\gamma$-rays generation method via single-shot laser plasma interaction. We find that in the weakly nonlinear regime of Compton scattering, the helicity of emitted photon is subjected to the interplay of the laser and electron helicities. Our proposed method can generate $\gamma$-photons with peak brilliance of $10^{21}$-$10^{22}~\mathrm{photons/(s \cdot mm^2 \cdot mrad^2 \cdot 0.1\% BW)}$ and average polarization degree of $\sim 21\%$-$37\%$ with moderate laser pulses. Moreover, our method is quite stable in a wide range of laser and plasma parameters. With proper selection of $\gamma$-photon energy and (polarized) plasma target,  highly CP brilliant $\gamma$-ray sources can be obtained for many applications, such as, photo-nuclear researches, generation of polarized lepton sources, etc.

\section*{Acknowledgements}
This work is supported by the National Natural Science Foundation of China (Grants Nos. 11905169, 12022506, 11874295), and the China Postdoctoral Science Foundation (Grant No. 2020M683447).

\bibliography{library}

\end{document}